\input harvmac

\Title{\vbox{\baselineskip12pt\hbox{UICHEP-TH/97-13}}}
{\vbox{\centerline{Lepto-mesons, Leptoquarkonium and the QCD Potential}}}

\centerline{David~Bowser-Chao\footnote{$^1$}{davechao@uic.edu},
Tom~D.~Imbo\footnote{$^2$}{imbo@uic.edu},
B.~Alex~King\footnote{$^3$}{aking@uic.edu} and
Eric~C.~Martell\footnote{$^4$}{ecm@uic.edu}}

\bigskip\centerline{Department of Physics}
\centerline{University of Illinois at Chicago}
\centerline{845 W.  Taylor St.}
\centerline{Chicago, IL \ 60607-7049}
\vskip 1.0in

We consider bound states of heavy leptoquark-antiquark pairs (lepto-mesons) as well as leptoquark-antileptoquark pairs (leptoquarkonium). Unlike the situation for top quarks, leptoquarks (if they exist) may live long enough for these hadrons to form. We study the spectra and decay widths of these states in the context of a nonrelativistic potential model which matches the recently calculated two-loop QCD potential at short distances to a successful phenomenological quarkonium potential at intermediate distances. We also compute the expected number of events for these states at future colliders.

\Date{}

Particle physicists often lament the fact that the top quark is so short-lived.  A long-lived top quark would have allowed us to study various aspects of the Standard Model (SM) that are otherwise inaccessible.  For instance, had top quarks lived long enough to form bound states, they would have provided an exciting window into the nature of the strong interactions. But the SM top quark width, $\Gamma_{t\rightarrow Wb}$, grows as $m_t^3$ for large $m_t$, and for $m_t=175$~GeV is about 1.55~GeV.  This corresponds to a lifetime of approximately $4\times 10^{-25}$~s --- very short-lived indeed. However, there may still be a possibility of addressing these questions --- not with top quarks, but with leptoquarks ($LQ$s).  These hypothetical particles carrying both lepton and baryon number appear naturally in many extensions of the SM. Some $LQ$s may even be relatively light, with masses on the order of a few hundred GeV. (There has been a recent flurry of interest in these models due to data from 
HERA~\ref\hera{ H1 Collaboration, C.~Adloff {\it et al.}, Z. Phys. C {\bf 74} (1997) 191\semi ZEUS Collaboration, J.~Breitweg {\it et al.}, Z. Phys. C {\bf 74} (1997) 207.} 
which initially seemed to hint at the existence of a 200~GeV leptoquark, although new results from the Fermilab Tevatron 
\ref\tev{D0 Collaboration, B.~Abbott {\it et al.}, Phys. Rev. Lett. {\bf 79}
(1997) 4321\semi CDF Collaboration, F.~Abe {\it et al.}, Phys. Rev. Lett. {\bf 79} (1997) 4327.} make this scenario unlikely.)
In what follows, we will restrict ourselves to models in which each leptoquark has at least one trilinear coupling with some ordinary quark-lepton pair.\foot{It is also usually assumed that a given leptoquark can only couple to quarks and leptons from a single generation --- hence the terminology {\it first generation leptoquark}, etc. We instead make the less restrictive assumption that a given leptoquark couples to quarks from only one generation and leptons from only one generation, but the two need not be the same.}  Such leptoquarks will be color triplets having a spin of either zero or one. Their widths, which only grow linearly with their masses, will be proportional to the square of the above unknown coupling strengths. These couplings cannot be too large otherwise leptoquarks would have already been seen.  Indeed, for reasonably large masses and small couplings, leptoquarks will have widths between about 1 and 100~MeV. As we will see below, this leaves plenty of room for the formation and observation of leptoquark bound states, either with another leptoquark (leptoquarkonium, $LQ$-$\overline{LQ}$) or with a $u$, $d$, $s$, $c$ or $b$ quark (lepto-mesons, \hbox{$LQ$-${\bar q}$}).\foot{Previous mention of bound states containing leptoquarks can be found in
\ref\lh{J.~Bl\" umlein, DESY preprint 93-153, contribution to the Proceedings of the Workshop ``$e^+e^-$ Collisions at 500 GeV,'' Munich, Annecy, Hamburg (1993)\semi C.~Friberg, E.~Norrbin and T.~Sj\"ostrand, Phys. Lett. B {\bf 403} (1997) 329\semi J.~L.~Hewett and T.~G.~Rizzo, hep-ph/9703337, 
version 3.}\ref\kis{V.~V.~Kiselev, hep-ph/9710432.}.} It may even be  reasonable to consider $LQ$-${\bar t}$ states.

The first step in identifying signals of these bound states is understanding their spectra. As with quarkonia, this is most easily done in a nonrelativistic potential model approach.  Since leptoquarks are color triplets, the static QCD potential between an $LQ$ and an $\overline{LQ}$, or between an $LQ$ and a ${\bar q}$, is the same as that between a $q$ and a ${\bar q}$ (although since leptoquarks have integral spin, relativistic corrections such as hyperfine splittings will be different). However, the region of the potential probed by our new bound states will be somewhat different than that probed by the known quarkonium states. Thus, we cannot blindly use a generic phenomenological potential which is tailored only to the latter, but must attempt to construct one more suited to our needs.  Of course, this potential should be consistent with the known results from QCD as well as  what we have learned from quarkonium studies. We will construct such a potential below (in the spirit of \ref\hag{K.~Hagiwara, A.~D.~Martin and A.~W.~Peacock, Z. Phys. C {\bf 33} (1986) 135.}) and then use it to evaluate the energy splitting and the square of the wavefunction at the origin, $|\psi (0)|^2$, for the $1S$ and $2S$ states of leptoquarkonium and lepto-mesons.  This will allow us to calculate approximate widths and production cross sections, and estimate the expected number of events (on resonance) for these states.

What do we know about the global structure of the static QCD potential $V(r)$? At ``short distances'' it is quasi-Coulombic, with a perturbative expansion in the running coupling $\alpha_s$. The ``long distance'' behavior, describing quarks bound by tubes of chromo-electric flux, is expected to be linear. There is currently no known analytic method to describe $V(r)$ between these regimes. However, many phenomenological potential models have been developed to fit the $c{\bar c}$ and $b{\bar b}$ spectra which basically probe this region, all of which have a quasi-logarithmic form at these distance scales.  Finally, we note that $V(r)$ is known to be concave downward for all $r$; that is, $V'(r)>0$ and $V''(r)<0$ 
\ref\r?{C. Borgs and E. Seiler, Comm. Math. Phys. {\bf 91} (1983) 329\semi C.~Bachas, Phys. Rev. D {\bf 33} (1986) 2723.}.

We first discuss the short range potential $V_{\rm short} \left( r \right)=-{4 \over 3} {\alpha_{\scriptscriptstyle V}\left(\mu\right) \over r}$,
where $\alpha_{\scriptscriptstyle V}$ can be expanded in terms of $\alpha_s$.  Until recently, this expansion was only known to one loop in the presence of $n_f$ massless flavors of quarks 
\ref\bil{A.~Billoire, Phys. Lett. {\bf 92B} (1980) 343.}, 
although it was known to two loops in the absence of fermions 
\ref\fis{W.~Fischler, Nucl. Phys. {\bf B129} (1977) 157.}.  
However, the {\it full} two loop potential has recently been computed 
\ref\peter{M. Peter, Nucl. Phys. {\bf B501} (1997) 471\semi M. Peter, Phys. Rev. Lett. {\bf 78} (1997) 602.}. 
The result, of course, is dependent on both the renormalization scale ($\mu =\mu (r)$) and scheme.  We will use the BLM method
\ref\r?{S.~J. Brodsky, G.~P. Lepage and P.~B. Mackenzie, Phys. Rev. D {\bf 28}
(1983) 228\semi S.~J.~Brodsky and H.~Lu, Phys. Rev. D {\bf 51} (1995) 3652.} 
for which
\eqn\blmpot{
V_{\rm short} \left( r \right)=-{16 \pi \over 3r} \left( \left({\alpha_{\overline{ \rm \scriptscriptstyle MS}} \left( \mu^{*} \right) \over 4 \pi }\right) + c_1  \left({\alpha_{\overline{\rm \scriptscriptstyle MS}} \left( \mu^{**} \right) \over 4 \pi }\right)^2 + c_2 \left({\alpha_{\overline{ \rm \scriptscriptstyle MS}} \left( \mu^{***} \right) \over 4 \pi}\right)^3 +\dots\ \right).
}
Here the computation is performed in the ${\overline {\rm MS}}$ scheme, and the scale at each order is chosen so that the coefficients $c_i$ are $n_f$-independent. Combining this with the general results of \peter\ yields $c_1=-8$ and $c_2=14/3 + 54 \pi^2 - 9 \pi^4/4 - 220 \zeta_3$, as well as 
\eqn\muone{
\mu^{*} \left( r \right)={1 \over r} e^{\textstyle -\left[\gamma + {5 \over 6} - {\pi ^2 \beta_0 \over 6} \left({\alpha_{\overline{ \rm \scriptscriptstyle MS}} \left( {\tilde \mu}\right) \over 4 \pi}\right)\right]};\qquad \mu^{**} \left( r \right)={1 \over r} e^{\textstyle -\left({49-52 \zeta_3 \over 16}\right)}
}
for the first two BLM scales (where $\gamma\simeq 0.57722$ and $\zeta_3 \simeq 1.20206$). To two loops, the BLM method does not determine the scales $\mu^{***}$ in \blmpot\ and ${\tilde \mu}$ in \muone. For simplicity, we choose both of these scales to be $\mu^{**}$.  An approximate value for $\alpha_{\overline{\rm \scriptscriptstyle MS}}$ at any scale $\mu$ can be found from a reference value $\alpha_{\overline{\rm \scriptscriptstyle MS}}\left( q \right)$ (we use $\alpha_{\overline{ \rm \scriptscriptstyle MS}}(m_{\scriptscriptstyle Z})=0.118$) and the following formula~\peter  
\eqn\alphams{
\eqalign{
\alpha_{\overline{\rm \scriptscriptstyle MS}} \left( \mu \right)= &\alpha_{\overline{\rm \scriptscriptstyle MS}} \left( q \right) \left[ 1 + \left({\alpha_{\overline{\rm \scriptscriptstyle MS}} \left( q \right) \over 4 \pi}\right) \beta_0 \ln \left( {\mu^2 \over q^2} \right) + \left({\alpha_{\overline{\rm \scriptscriptstyle MS}} \left( q \right) \over 4 \pi}\right)^2 \beta_1 \ln \left( {\mu^2 \over q^2} \right) \right. \cr
&\qquad \left.+ \left({\alpha_{\overline{\rm \scriptscriptstyle MS}} \left( q \right) \over 4 \pi}\right)^3 \left( \beta_2 \ln \left( {\mu^2 \over q^2} \right) -{1 \over 2} \beta_0 \beta_1 \ln^2 \left( {\mu^2 \over q^2} \right) \right)\right]^{-1},
}}
which, like the potential, is complete to third order. Here
\eqn\betas{{
\beta_0=11 - {2\over 3}n_f;\qquad
\beta_1=102 - {38\over 3}n_f;\qquad
\beta_2={2857 \over 2} - {5033\over 18}n_f + {325\over 54}n_f^2.
}}
This form of the potential is only valid at energy scales far above quark pair creation thresholds, where all $n_f$ quark flavors can be considered ``light''. Since this will be far from true in the applications that follow, we need a way to incorporate the effect of quark masses. A naive method would be to make $n_f$ a discontinuous function of $\mu$, increasing by one unit as each threshold is crossed. However, this is not very realistic. In the absence of a full two loop calculation with massive quarks, we shall instead model these effects in the manner suggested in
\ref\nf{S.~J.~Brodsky, M.~S.~Gill, M.~Melles and J.~Rathsman, hep-ph/9801330.}, letting $n_f$ be an analytic function of $\mu$:
\hbox{$n_f(\mu )=\sum_q {\mu^2\over(\mu^2+{\widetilde m}_q^2)}.$}
Here, the sum is over all quark flavors and the ${\widetilde m}_q$'s are the current quark masses (not to be confused with the constituent masses used later). As in \nf\ we choose ${\widetilde m}_u=0.004$~GeV, ${\widetilde m}_d=0.008$~GeV, ${\widetilde m}_s=0.2$~GeV, ${\widetilde m}_c=1.5$~GeV, ${\widetilde m}_b=4.5$~GeV and ${\widetilde m}_t=175$ GeV. (Our final results are rather insensitive to these numerical choices.) We believe this definition of $V_{\rm short}(r)$ to be valid out to $r\simeq 0.3~{\rm GeV^{-1}}$.

The ``intermediate distance'' range is the least understood (from first principles) portion of the QCD potential.  Running from the edge of the perturbative short distance region out to the long distance linear regime, it includes the quasi-logarithmic range (\hbox{$r\simeq0.9$--$5$~GeV$^{-1}$}) where the low-lying $c{\bar c}$ and $b{\bar b}$ states live. Since the properties of certain bound states containing leptoquarks will be sensitive to the transition between the perturbative and logarithmic regions, we want to connect them as smoothly as possible. To this end we would like an intermediate potential which attaches to our perturbative potential at some reasonable distance scale $r_a$ in such a way that $V$, $V'$ and $V''$ are continuous at $r=r_a$. We also require that $V(r)$ does a reasonable job of fitting the known (spin-averaged) $c{\bar c}$ ($1S$, $2S$ and $1P$) and $b{\bar b}$ ($1S$, $2S$, $3S$, $1P$ and $2P$) states.  We choose to match at $r_a=0.2$ GeV$^{-1}$, which is safely (but not too far) within the perturbative region. Since the first two derivatives of $V$ will be continuous here, our results are somewhat insensitive to this choice. We choose the form of the intermediate potential to be a slightly modified version of the standard log potential 
\ref\logref{C.~Quigg and J.~Rosner, Phys. Lett. {\bf 71B} (1977) 153.}:
\eqn\log{
V_{\rm int}\left(r\right) = A \ln\left({r\over r_0}\right) + a e^{-b\left( r-r_a \right)}.
}
As in \logref\ we take $A=0.733$~GeV. The constants $r_0$, $a$ and $b$ are determined by our continuity criteria to be $r_0\simeq 5.501$ GeV$^{-1}$, $a\simeq -0.013$ GeV and  $b\simeq 49.30$ GeV. The second term in \log\ is non-negligible only in the region $r\simeq 0.2$--$0.4$ GeV$^{-1}$, so we do not expect it to have much of an effect on the quarkonium states. 
\vskip -36pt
\vbox{\offinterlineskip
\vbox{
\let\picnaturalsize=Y
\def\picsize{1.0in}
\def\picfilename{QCDpot.eps}
\ifx\nopictures Y\else{\ifx\epsfloaded Y\else\input epsf \fi
\global\let\epsfloaded=Y
\centerline{\ifx\picnaturalsize N\epsfxsize \picsize\fi \epsfbox{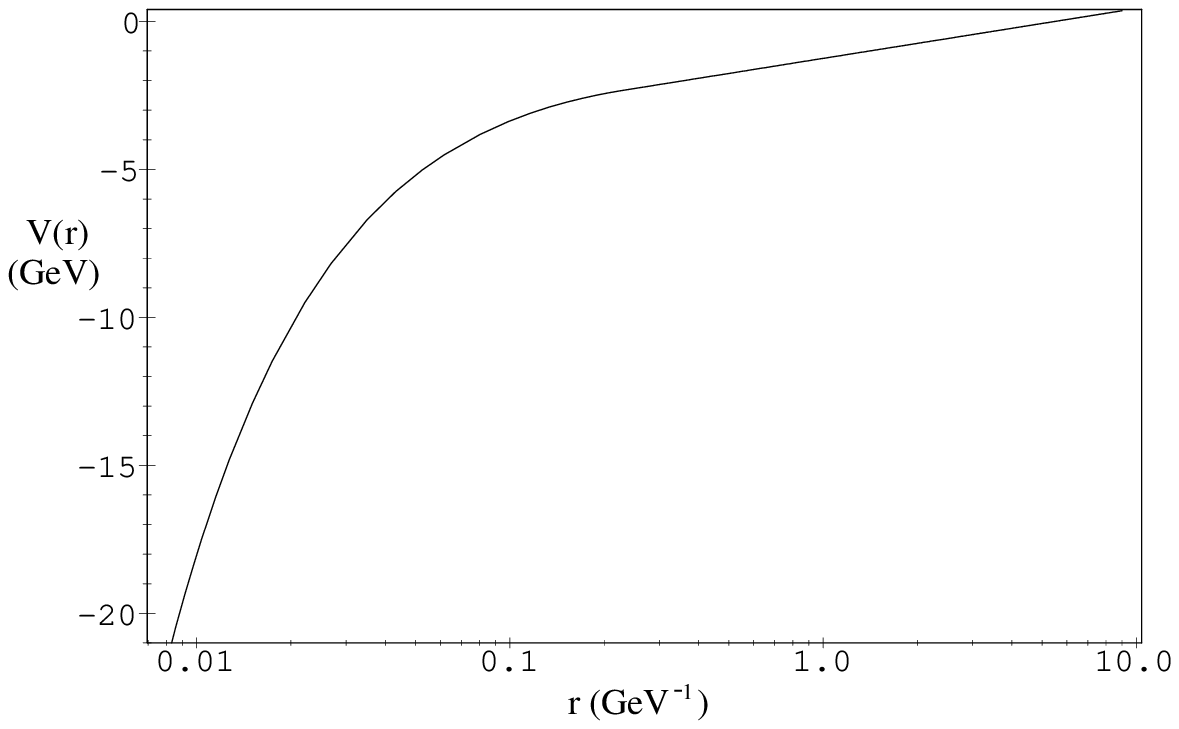}}}\fi
}
\vskip -22pt
\vbox{
\centerline{{\bf Figure 1:}  Our version of the static QCD potential.}
}}\vskip 12pt

As it stands, our potential does not have the appropriate linear behavior at ``large'' distances. We can, of course, fix this in any number of ways. However, we feel that this is unnecessary for our purposes since most of the states in which we are interested are not sensitive to this linear regime. The few states that we consider which have somewhat small reduced masses ($\,${\lower-1.2pt\vbox{\hbox{\rlap{$<$}\lower5pt\vbox{\hbox{$\sim$}}}}}$\,$0.5~GeV) still lie mostly in the region well-described by the log potential (at least in the $1S$ case --- see, for example, \ref\light{W.~Kwong and J.~L.~Rosner, Phys. Rev. D {\bf 44} (1991) 212\semi T.~D.~Imbo, Phys. Lett. B {\bf 398} (1997) 374.}).  In any event, the whole nonrelativistic approach of this paper is suspect for these states. We only include them in order to show certain qualitative trends.\foot{We have also ignored the effects of leptoquark-Higgs interactions in our potential. These very model-dependent couplings, even if large, would only affect the $LQ$-$\overline{LQ}$ (or possibly $LQ$-${\bar t}$) states --- and even then only if the Higgs mass is ``small enough''. We neglect the model-dependent leptoquark couplings to electroweak gauge bosons as well.} Our full potential (Figure~1) is analytic except at $r=r_a$, and is everywhere concave downward. 

Using the above potential and the nonrelativistic Schr\"odinger equation, we have numerically computed the bound state energies and wavefunctions at the origin for the $1S$ and $2S$ states of lepto-mesons and leptoquarkonium (Table 1), for various constituent leptoquark masses between 300~GeV and 1~TeV.  (We also checked that $V(r)$ fits the low-lying $c{\bar c}$ and $b{\bar b}$ states with approximately the same $\chi^2$ as the standard log potential alone.)
\vskip 12pt
\vbox{\offinterlineskip
$$\vbox{
\halign{\strut\vrule\quad\hfil#\hfil\quad&\vrule\quad\hfil#\hfil\quad&\vrule\quad\hfil#\quad
&\vrule\quad\hfil#\quad&\vrule\quad\hfil#\hfil\quad\vrule\cr
\noalign{\hrule}
$m_{\scriptscriptstyle LQ}$&System&$|\psi_{1S}(0)|^2$\hfil&$|\psi_{2S}(0)|^2$\hfil&$\Delta E_{2S-1S}$\cr
\noalign{\hrule}
300&$LQ$-${\bar u}$/${\bar d}$&0.020&0.010&0.589\cr
&$LQ$-${\bar s}$&0.035&0.018&0.589\cr
&$LQ$-${\bar c}$&0.190&0.097&0.589\cr
&$LQ$-${\bar b}$&1.285&0.649&0.593\cr
&$LQ$-${\bar t}$&1,985&308.5&1.346\cr
&$LQ$-$\overline{LQ}$&4,574&640.8&1.751\cr
\noalign{\hrule}
500&$LQ$-${\bar c}$&0.191&0.097&0.589\cr
&$LQ$-${\bar b}$&1.299&0.656&0.593\cr
&$LQ$-${\bar t}$&3,072&445.8&1.542\cr
&$LQ$-$\overline{LQ}$&1.819$\times10^4$&2,622&2.725\cr
\noalign{\hrule}
700&$LQ$-${\bar c}$&0.191&0.097&0.589\cr
&$LQ$-${\bar b}$&1.305&0.659&0.593\cr
&$LQ$-${\bar t}$&3,791&538.1&1.649\cr
&$LQ$-$\overline{LQ}$&4.489$\times 10^4$&6,776&3.604\cr
\noalign{\hrule}
1000&$LQ$-${\bar c}$&0.191&0.097&0.589\cr
&$LQ$-${\bar b}$&1.310&0.662&0.593\cr
&$LQ$-${\bar t}$&4,486&629.2&1.741\cr
&$LQ$-$\overline{LQ}$&1.167$\times 10^5$&1.820$\times 10^4$&4.808\cr
\noalign{\hrule}
}}$$
\vskip 6pt
\centerline{\vbox{\hsize=10cm\noindent
{\bf Table 1:} Values of $|\psi_{nS}(0)|^2$ and the $1S$-$2S$ energy splitting for lepto-mesons and leptoquarkonium. All quantities are in appropriate GeV units.}}}

\noindent
This analysis also required choices for the constituent quark masses. For the $c$ and $b$ quarks we used those from \logref : $m_c=1.5$~GeV and $m_b=4.906$~GeV. For the top quark we used $m_t=175$~GeV, while for the lighter quarks we chose $m_s=0.5$ GeV and \hbox{$m_u=m_d=0.35$~GeV}. The states $LQ$-${\bar u}/{\bar d}$ and $LQ$-${\bar s}$ are only presented for a leptoquark mass of 300~GeV, since the reduced mass of the system is basically independent of the leptoquark mass in the range of interest. The radii of the $1S$ heavy-heavy states in Table 1 ($LQ$-$\overline{LQ}$ and $LQ$-${\bar t}$) lie well within the perturbative regime of $V(r)$, ranging from approximately  0.02 to 0.05~GeV$^{-1}$ as the reduced mass decreases. The $2S$ heavy-heavy states have radii from about 0.05 to 0.22~GeV$^{-1}$. Here we begin to see some sensitivity to the short-intermediate matching region for lighter leptoquark masses. The $1S$ $LQ$-${\bar b}$ states all lie around 0.6~GeV$^{-1}$, and are also somewhat sensitive to the matching. All other states considered have radii of 1.2 GeV$^{-1}$ and above, and thus mainly see the logarithmic part of the potential. (For an alternative treatment of $LQ$-${\bar c}$ and $LQ$-${\bar b}$ states, see \kis .)  
 
Technically, the results in Table 1 represent {\it spin-averaged} quantities since we have not included hyperfine effects in our potential. However, for scalar leptoquarks this statement is trivial since their leptoquarkonium states can only have spin-0 and their lepto-meson states spin-1/2. For vector leptoquarks there {\it are} hyperfine splittings.  These are negligible for the lepto-mesons containing lighter quarks, but larger for the $LQ$-${\bar t}$ states and for leptoquarkonium.  However, they are still substantially smaller than the $1S$-$2S$ splittings (\hbox{{\ \lower-1.2pt\vbox{\hbox{\rlap{$<$}\lower5pt\vbox{\hbox{$\sim$}}}}\ }0.1~$\Delta E_{2S-1S}$}), justifying our assumption that the static potential alone is a good starting point. For simplicity, in the remainder of this paper we will only consider scalar leptoquarks.

Our analysis so far has only utilized the mass and spin of the leptoquark, along with the assumption that it transforms as an $SU(3)_c$ triplet. The study of leptoquark bound state production and decay, however, is more model-dependent; for example, $\gamma\gamma \to LQ$-$\overline{LQ}$ involves the leptoquark's electric charge. We  follow
\ref\buch{W.~Buchm\"uller, R.~R\"uckl and D.~Wyler, Phys.~Lett.~{\bf B191} (1987) 442.} 
in assuming that at the mass scales considered, the SM is extended only by the addition of leptoquarks which respect baryon and lepton number and $SU(3)_c \times SU(2)_L \times U(1)_Y$. The leptoquark isospin and hypercharge then determine its coupling to the SM gauge bosons, and to which quark-lepton pairs it may couple. We explicitly consider only one (presumably the lightest) scalar leptoquark. A degenerate weak isospin doublet or triplet of leptoquarks would not drastically change our conclusions, but could enrich the phenomenology (by giving rise to the decay \hbox{$LQ$-$\overline{LQ} \to W^+ W^-$}, for example). Except in the case of $\ell^+ \ell^- \to LQ$-$\overline{LQ}$, we assume a purely chiral leptoquark coupling; if its trilinear coupling to a lepton of chirality $i$ ($i=L,R$) and quark of opposite chirality is denoted by $\lambda_i$, then we assume that either $\lambda_L\ne 0, \lambda_R = 0$ or $\lambda_R\ne 0, \lambda_L = 0$ in order to satisfy magnetic moment constraints (see below). For convenience, we write $\lambda_i = e\tilde\lambda_i$, with $e^2/4\pi=\alpha_{em}(m_{\scriptscriptstyle Z})=1/128$. To ensure that the bound state widths are substantially smaller than the $1S$-$2S$ splitting (making the states easily identifiable), we require weakly coupled leptoquarks ($\tilde\lambda_i^2  {\ \lower-1.2pt\vbox{\hbox{\rlap{$<$}\lower5pt\vbox{\hbox{$\sim$}}}}\ } 1$). This leads to bound state widths of at most  ${\cal O}(100\; \hbox{MeV}$). Finally, we continue to ignore the model-dependent leptoquark couplings to $W$, $Z$ and Higgs bosons --- including the effects of partial widths such as \hbox{$LQ$-$\overline{LQ}\to ZZ$} would not greatly change our results. 

The production cross-section (on resonance) of a narrow width bound state $B$ from an initial state $i$, summed over all decay modes of the resonance 
\ref\resonance{See, for example, S. Weinberg, Quantum Theory of Fields, Vol. I, (Cambridge 1995) Eq. 3.8.16.}, is well-approximated by \hbox{$\sigma_{\hbox{tot}} = {\left(16 \pi S/M_B^2\right) \left(\Gamma_{B\to i} \over\Gamma\right)}$}. Here, $\Gamma$ and $\Gamma_{B\to i}$ are the total and partial widths of the bound state, and the rest-frame total energy $E$ is well within $\Gamma$ of the resonance mass $M_B$. In the particular case of unpolarized incoming beams, the spin-multiplicity factor $S$ is given by $N_B/(N_1 N_2)$, with $N_B$ the spin-multiplicity of the resonance and $N_1,N_2$  those of the incoming particles comprising the state $i$. When the colliding beams are not monochromatic compared to $\Gamma$, we assume a gaussian distribution in $E$ strongly peaked at the resonance mass $M_B$ with a spread of $\sigma_{\scriptscriptstyle E}$ ($\sigma_{\scriptscriptstyle E}${\ \lower-1.2pt\vbox{\hbox{\rlap{$>$}\lower5pt\vbox{\hbox{$\sim$}}}}\ }$\Gamma$). At the energies considered, \hbox{$\sigma_{\hbox{tot}} = {\left(16\pi S/M_B^2\right)\left( \sqrt{\pi\over8}\right)\left({\Gamma_{B\to i}\over\sigma_{\scriptscriptstyle E}}\right)}$}.  We assume that leptoquarks would first be discovered in non-bound state production, and that $m_{\scriptscriptstyle LQ}$ would be sufficiently well-determined to allow a search for the resonances considered here.

We first treat lepto-mesons ($M_B\simeq m_{\scriptscriptstyle LQ}+m_q$). Throughout, our notation will not distinguish between particle and antiparticle so that (for example) $LQ$-$q$ represents \hbox{$LQ$-${\bar q}$} or $\overline{LQ}$-$q$, and $\Gamma_{LQ\to eq}$ represents $\Gamma_{LQ\to e^{\pm}q}$ or $\Gamma_{\overline{LQ}\to e^{\pm}\overline{q}}$. 
The principal lepto-meson
decay channels should be the spectator decays of the quark ($\Gamma_{B\to (LQ)q^{'}+X} = \Gamma_{q\to q^{'}+X}$) and leptoquark:
\eqn\lqdecay{\Gamma_{B\to\ell qq} = \Gamma_{LQ\to\ell q} = {\alpha_{em}
\tilde\lambda_i^2 \over 4} \left({m_{\scriptscriptstyle LQ}^2-m_q^2 \over
m_{\scriptscriptstyle LQ}^2}\right)^2 m_{\scriptscriptstyle LQ} \hskip 0.25in (\ell=e,\mu ,\tau).}
(We will ignore leptoquark couplings to neutrinos as well as all lepton masses.)
As mentioned earlier, for large $m_{\scriptscriptstyle LQ}$ the leptoquark decay
width increases only as $m_{\scriptscriptstyle LQ}$.
The quark decay width $\Gamma_{B\to (LQ)q^{'}+X}$
is negligible except in the case of the top quark, where 
$\Gamma_{B\to (LQ)q^{'}+X}
=\Gamma_{t \rightarrow W b}$. We will also need the
very narrow width $\Gamma_{B\to\ell_a\gamma_b} = {(1+ab)\over 4}\Gamma_{B\to \ell\gamma}$ for decay into $\ell\gamma$ with helicities $a/2$ and
$b$ respectively:
\eqn\begamun{\Gamma_{B\to\ell\gamma} = {N_c\over 16\pi}
\left(4\pi\alpha_{em}\tilde\lambda_i\right)^2 \left|\psi(0)\right|^2 \left({Q_{\ell}\over
M_B}-{Q_q\over m_q}\right)^2 {M_B\over m_{\scriptscriptstyle LQ}}.}
Here $N_c=3$ and $Q$ is the electric charge (in units of the proton charge).
The width $\Gamma_{B\to\ell\gamma}$ 
is inconsequential as a contribution to the total width, which
is well-approximated by \hbox{$\Gamma = \Gamma_{B\to (LQ)q^{'}+X} + \Gamma_{B\to\ell qq}$.} 
Over the mass range we consider, the $LQ$-$t$ width $\Gamma_{B\to\ell tt}$ ranges from 
\hbox{$\tilde\lambda_i^2\cdot 255$~MeV} to 
\hbox{$\tilde\lambda_i^2\cdot 1835$~MeV},
while for lighter quarks the $LQ$-$q$ widths 
$\Gamma_{B\to\ell qq}$ all range from about 
\hbox{$\tilde\lambda_i^2\cdot 586$~MeV} to \hbox{$\tilde\lambda_i^2\cdot 1953$~MeV}. We may now consider the production of lepto-mesons at an $e^-\gamma$ collider (for non-bound state production, see
\ref\nadeau{H.~Nadeau and D.~London, 
Phys. Rev. D {\bf 47} (1993) 3742.}). With an
expected beam spread of $\sigma_{\scriptscriptstyle E}=0.05\sqrt s\approx
0.05M_B$ \ref\gamgam{R.~Brinkmann, I.~Ginzburg, N.~Holtkamp, G.~Jikia,
O.~Napoly, E.~Saldin, E.~Schneidmiller, V.~Serob, G.~Silvestrov,
V.~Telnov, A.~Undrus and M.~Yurkov, Nucl. Inst. Meth. A {\bf 406} (1998) 13.}, the cross-section for a polarized $e^-\gamma$
collider is \hbox{$\sigma_{\hbox{tot}, ab}={\left(16\pi/M_B^2\right)\left(\sqrt{\pi\over 8}\right)\left(\Gamma_{B\to{e_a\gamma_b}}\over\sigma_{\scriptscriptstyle E}\right)}$}. The spin-multiplicity factor $S$ has been specified implicitly in
this case ($S=1$) as is done below for the other  cross-sections
computed. For unpolarized incoming beams, $S=1/2$. 

With the exception of $LQ$-$t$, for which the
top decay width dominates, the $1S$-$2S$ splitting is larger than the width of the states for $\tilde\lambda_i^2 {\ \lower-1.2pt\vbox{\hbox{\rlap{$<$}\lower5pt\vbox{\hbox{$\sim$}}}}\ } 1/10$.
(For $LQ$-$t$, there may still be the possibility of resolving the $1S$-$2S$ peaks for very heavy leptoquarks --- see Table~1.)
In Table 2 we present the expected number of events per $\tilde\lambda^2_i$ for the production of $1S$ states.  Given a yearly
integrated luminosity of up to 50 $\hbox{fb}^{-1}$\gamgam, these resonances
could be observable.
Resolving these states, however, will be difficult in practice
given the broad collider energy distribution and the non-trivial task of
precisely reconstructing the lepto-meson given that there is a jet in the final state. 
\vskip 12pt
\vbox{\offinterlineskip
$$\vbox{
\halign{\strut\vrule\quad\hfil#\hfil\quad
&\vrule\quad\hfil#\hfil\quad
&\vrule\quad\hfil#\hfil\quad
&\vrule\quad\hfil#\hfil\quad
&\vrule\quad\hfil#\hfil\quad
&\vrule\quad\hfil#\hfil\quad
&\vrule\quad\hfil#\hfil\quad\vrule\cr
\noalign{\hrule}
\vrule height11pt depth3.5pt width0pt $m_{\scriptscriptstyle LQ}$&$ \overline{LQ}\hbox{-}t$&$\overline{LQ}\hbox{-}b$&$\overline{LQ}\hbox{-}c$&$\overline{LQ}\hbox{-}s$&
$\overline{LQ}\hbox{-}u$&$\overline{LQ}\hbox{-}d$\cr
\noalign{\hrule}
300&72.36&13.58&98.51&40.36&194.5&48.11\cr
\noalign{\hrule}
500&26.63&3.127&21.31& 8.765&41.98&10.43\cr
\noalign{\hrule}
700&12.24&1.171&7.769&3.202&15.30&3.807\cr
\noalign{\hrule}
1000&4.979&0.410&2.665&1.100&5.246&1.307\cr
\noalign{\hrule}
}}$$
\vskip 6pt
\centerline{\vbox{\hsize=10cm \noindent
{\bf Table 2:} Estimates of the number of events/$\tilde\lambda_i^2$ for $1S$ lepto-mesons produced approximately on resonance at a polarized $e^-\gamma$ collider.  Masses are in GeV. We assume $\sigma_{\scriptscriptstyle E}=0.05M_B$ and an integrated luminosity of $10$ fb$^{-1}$. The particle-antiparticle assignments shown in the column headings were chosen to maximize production.
}}}
\vskip 6pt
Turning to leptoquarkonium ($M_B\simeq 2m_{\scriptscriptstyle LQ}$), the partial width for the spectator decay of either constituent leptoquark is 
$\Gamma_{B\to\ell q(LQ)} = 2 \Gamma_{LQ\to\ell q}$.
For a very weak trilinear coupling ($\tilde\lambda_i^2  {\ \lower-1.2pt\vbox{\hbox{\rlap{$<$}\lower5pt\vbox{\hbox{$\sim$}}}}\ } 1/1000$) this decay channel is not dominant. The two gluon or two photon (with helicities $a$ and $b$) decay modes are more significant:
\eqn\glueglue{\Gamma_{B\to g_{a}g_{b}} = {\delta_{ab} \over 2\pi N_c} { \left|\psi(0)\right|^2 \over M_B^2} \left(4\pi\alpha_s\right)^2 ,}
\eqn\ppunpol{\Gamma_{B\to \gamma_{a}\gamma_{b}} = {\delta_{ab} \over 2} \Gamma_{B\to \gamma\gamma};\qquad\Gamma_{B\to\gamma\gamma} =
	{1\over 2\pi }N_c { \left|\psi(0)\right|^2 \over M_B^2}
\left(4\pi\alpha_{em}\right)^2 Q_{LQ}^4.}
On the other hand, for $\tilde\lambda_i^2 {\ \lower-1.2pt\vbox{\hbox{\rlap{$>$}\lower5pt\vbox{\hbox{$\sim$}}}}\ }1/100$ we have $\Gamma \simeq \Gamma_{B\to\ell q(LQ)}$.
We can now discuss the production of leptoquarkonium at a $\gamma\gamma$ collider. Assuming a $\gamma\gamma$ center of mass energy resolution $\sigma_{\scriptscriptstyle E}$ of $0.15\sqrt s\approx 0.15 M_B$ \gamgam , the cross-section for unpolarized incoming photons is
\eqn\sigtotgamgam{
\sigma_{\hbox{tot}} = {16\pi\over M_B^2}\sqrt{\pi\over 8}{\Gamma_{B\to
\gamma\gamma} \over 2\sigma_{\scriptscriptstyle E}}.
}
(There is an additional factor of 2 in this equation to compensate for the factor of 1/2 that was included in Eq.~\ppunpol\ to account for the identical photons in the final state.)
Eq.~\ppunpol~shows that production will be greatly enhanced or suppressed
depending on the value of
$Q_{LQ}~(\left|Q_{LQ}\right| = 1/3, 2/3, 4/3~{\rm or}~5/3)$. We have taken the most optimistic possibility, $\left|Q_{LQ}\right|= 5/3$, in \hbox{Table 3}. It is interesting to consider a very small $\tilde\lambda_i^2$, say $\tilde\lambda_i^2 =1/10000$, where the dominant decay width ($\sim 1.7$~MeV for $m_{\scriptscriptstyle LQ}=300$~GeV) is into gluon pairs. In this
case the $\gamma\gamma$ decay channel, with a partial width of 0.45~MeV, has a branching ratio of about 1/5. The resulting narrow resonance in
$\gamma\gamma\to\gamma\gamma$ will stand out in a spectacular way over the
continuum background \ref\gamback{G.~Jikia
and A.~Tkabladze, Phys. Lett.  B {\bf 323} (1994) 453.}, which should be
more than an order of  magnitude smaller (at somewhat less than 1 fb within a
reasonable mass bin). Similar results are obtained even for $\left|Q_{LQ}\right|$ as
small as $2/3$.

Finally, for leptoquarkonium production at an $e^+e^-$ or $\mu^+\mu^-$  collider
we will need the decay width into a pair of charged leptons $(\ell^+\ell^-)$ 
with helicities $a/2$ and $b/2$:
\eqn\ee{
\Gamma_{B\to \ell^{+}_{a}\ell^{-}_{b}} =  \delta_{ab} 32\pi N_c \left(\alpha_{em}
{\tilde\lambda_{+a}}{\tilde\lambda_{-a}}\right)^2  {\left|\psi(0)\right|^2 m_q^2
\over(M_B^2+4m_q^2)^2}.
}
(For a review of non-bound state production, see \ref\bergerl{M.~S.~Berger, hep-ph/9609517.}.)
The cross-section for a polarized $\ell^{+}\ell^{-}$ collider operating at the
resonance is then \hbox{$\sigma_{\hbox{tot},ab} = (16\pi/M_B^2) \left(\Gamma_{B\to \ell^+_a\ell^-_b}\over \Gamma \right)$}, which vanishes unless we relax our condition that the coupling be purely chiral; in fact, it is maximized if we take $\tilde\lambda_L=\tilde\lambda_R$ (as in Table~3). The maximal cross-section will be for an $LQ$ that couples to $\ell t$, which we assume here. Since the next generation $e^+e^-$ colliders are expected to attain at best $\sigma_{\scriptscriptstyle E}\simeq 0.01\sqrt{s}\gg\Gamma$, and the measured electron magnetic moment requires an extremely small value for $\tilde\lambda_L\tilde\lambda_R$ 
\ref\emu{A.~Djouadi, T.~K\" ohler, M.~Spira and J.~Tutas, Z. Phys. C {\bf 46} (1990) 679.} (and thus a tiny partial width $\Gamma_{B \to e^+e^-}$),  production at these machines would be heavily suppressed. By contrast, proposed muon colliders could attain
$\sigma_{\scriptscriptstyle E} \simeq 2.1 \times 10^{-5} \sqrt{s}$ \ref\demarteau{M. Demarteau and T. Han,
hep-ph/9801407.}. Furthermore, the constraints from the muon $g-2$ are much looser,
allowing $\tilde\lambda_L^2=\tilde\lambda_R^2$ as large as 0.14 for $m_{\scriptscriptstyle LQ} \ge 300$~GeV. Results for \hbox{$\mu^+ \mu^- \to LQ$-$\overline{LQ}$} may be found in Table~3.  In this case, the useful decay mode would most likely be $B\to \mu^- q\mu^+ {\bar q}$ (with a branching ratio of 0.97 for $m_{\scriptscriptstyle LQ}=300$~GeV) so that this channel would not only be visible, but might also have the potential of resolving the $1S$-$2S$ splitting.
\vskip 18pt
\vbox{\offinterlineskip
$$\vbox{
\halign{\strut\vrule\quad\hfil#\hfil\quad&\vrule\quad\hfil#\hfil\quad&
\vrule\quad\hfil#\hfil\quad\vrule\cr
\noalign{\hrule}
\vrule height11pt depth3.5pt width0pt $m_{\scriptscriptstyle LQ}$&$\mu^+\mu^- \rightarrow {LQ\hbox{-}\overline{LQ}}$&$\gamma \gamma \rightarrow {LQ\hbox{-}\overline{LQ}}$\cr
\noalign{\hrule}
300&578.0&853.8\cr
\noalign{\hrule}
500&52.55&264.0\cr
\noalign{\hrule}
700&12.05&121.1\cr
\noalign{\hrule}
1000&2.568&52.91\cr
\noalign{\hrule}
}}$$
\vskip 6pt
\centerline{\vbox{\hsize=10cm\noindent
{\bf Table 3:} Estimates of the number of events (on resonance) at polarized $\mu^+\mu^-$ and unpolarized $\gamma\gamma$ colliders for $1S$ leptoquarkonium. Masses are in GeV and the integrated luminosity is 10~fb$^{-1}$. In the $\mu^+\mu^-$ case, we assume that $\sigma_{\scriptscriptstyle E}<\Gamma$, $\left|Q_{LQ}\right|=1/3$, and $LQ$ couples to $\mu t$ with $\tilde\lambda_L^2=\tilde\lambda_R^2=0.1$. In the $\gamma\gamma$ case, we take $\sigma_{\scriptscriptstyle E}=0.15M_B$ and $\left|Q_{LQ}\right|=5/3$.
}}}
\vskip 12pt
In conclusion, we have constructed a version of the static QCD potential which matches (a slightly updated form of) the recently calculated two-loop perturbative piece at short distances to a successful phenomenological quarkonium potential
at intermediate distance scales.  We then used this potential to determine the spectra and decay widths of bound states containing heavy leptoquarks; namely, lepto-mesons and leptoquarkonium.  The expected number of events for these states at future colliders were also estimated.  For reasonable values of leptoquark masses and couplings, many of these exotic hadrons should be observable.
\vskip 12pt
\noindent
{\it Note:} After the completion of this work, we noticed that a similar
version of the static QCD potential has recently been constructed in
\ref\jez{M.~Jezabek, J.~H.~Kuhn, M.~Peter, Y.~Sumino and T.~Teubner,
hep-ph/9802373.}, although used for different purposes. 
\vskip 12pt
\goodbreak
\noindent 
\centerline{\bf Acknowledgements}
\vskip 6pt

We thank Markus Peter for clarification of his results.  We also thank Mark Adams, Russell Betts, Adam Falk, Ben Grinstein, Wai-Yee Keung, Young-Kee Kim, Julius Solomon and Uday Sukhatme for helpful discussions. This research was supported in part by the U.S. Department of Energy under Grant
Number DE-FG02-91ER40676.

\listrefs

\raggedbottom

\bye